# Magneto-transport and thermoelectric studies of antiperovskite semimetal: $Mn_3SnC$

**Sunil Gangwar, Sonika, C. S. Yadav**

School of Physical Sciences, Indian Institute of Technology Mandi, Mandi-175075 (H.P.) INDIA

We explore the magnetotransport and thermoelectric (Seebeck and Nernst coefficients) properties of $Mn_3SnC$, an antiperovskite magnetic Nodal line semimetal. $Mn_3SnC$ shows paramagnetic (PM) to concurrent antiferromagnetic (AFM)/ferromagnetic (FM) transition at $T \sim 286$ K. The electrical resistivity and Seebeck coefficient indicate the importance of electron–magnon scattering in the concurrent AFM/FM regime. We observed a large positive magnetoresistance (MR) of $\sim 8.2$ at 8 T field near magnetic transition, in the otherwise negative MR behaviour for low temperatures. The electrical resistivity and MR show a weak thermal hysteresis around the boundary of transition temperature and the width of hysteresis decreases as magnetic field increases. Interestingly the Hall and Seebeck coefficients change sign from positive to negative below the transition temperature, highlighting the different scattering for holes and electrons in this multi-band system. The Seebeck and Nernst signal exhibit two sharp anomalies; one at the transition temperature and another at $\sim 50$ K. The anomaly at magnetic transition in the Nernst signal disappear at 8 T magnetic field, owing to the reduction of magnetic fluctuation. A pseudo-gap near the Fermi level produces an upturn with a broad minimum in the Seebeck signal.

## INTRODUCTION

Antiperovskite materials has got renewed interested recently due to the observation of topological electronic states [1-5]. Among these $Sr_3SnO$, $Sr_3PbO$ and $Ca_3PbO$ are Dirac semimetals, while Mn based compounds $Mn_3ZnC$ is shown to be a nodal line semimetal with intriguing magnetic properties [1-4]. $Mn_3NiN$, $Mn_3CuN$ and $Mn_3Ni_{1-x}Cu_xC$ antiperovskite materials have non – trivial transport properties such as anomalous Hall effect and anomalous Nernst effect [6-8]. S. M. L. Teicher *et al* predicted that $Mn_3ZnC$ show nodal line semimetal properties with separate Weyl nodes near the Fermi level [4]. The antiperovskite family $Mn_3XC$ (X = Sn, Zn, Ga) has complex magnetic ordering, so interplay between magnetism and topology can promotes fascinating transport properties. $Mn_3SnC$ is particularly interesting because of complex magnetism and topological surface states." Moreover, antiperovskite compounds have been explored to use for photovoltaic devices, thermoelectric generators, energy storage devices and large magnetocaloric effect. Many antiperovskite thermoelectric materials $X_6SOA_2$ (X = Na, K and A = Cl, Br, I) and $AsPX_3$ (X = Mg, Ca, and Sr) are non-toxic and show a potential candidature for renewable energy devices, can convert waste heat to usual carbo-free electric energy. The antiperovskite materials are useful to cheaper energy device, which are eco-friendly and non-toxic in nature [9-12].

$Mn_3ZnC$ shows first order antiferromagnetic (AFM) to ferromagnetic (FM) transition at $T_t \sim 233$ K and second order FM to paramagnetic transition at $T_C \sim 380$ K [4, 13-14]. In $Mn_3ZnC$, the electronic distortion occurs due to the AFM ordering, which is responsible for spin population energy shifts and the formation of a pseudo-gap. In spite of significant pseudo-gap, $Mn_3ZnC$ has finite number of Weyl nodes near the Fermi level [4]. $Mn_3GaC$ shows AFM to FM transition at $T_t \sim 178$ K, followed by FM to PM transition at $T_c \sim 242$ K [14-16]. $Mn_3SnC$ show first order magnetic transition at transition temperature $T_C \sim 286$ K from paramagnet (PM) state to a magnetically complex order state [17-19]. Neutron diffraction studies suggest that there are two types of Mn atoms with different spin alignment, Mn1 and Mn2 located, at the face centers of the antiperovskite cube. Mn2 align in a square configuration with a net antiferromagnetic moment of 2.4 $\mu_B$/Mn in a plane perpendicular to the z-axis. Mn1 align ferromagnetically with a moment of 0.7 $\mu_B$/Mn as shown in the inset of inset figure (1) [18,20]. Mn based antiperovskite compounds also show magnetocaloric effects (MCE) around their magnetic transition, with entropy change $\Delta S \sim 133$ mJ $cm^3$ $K^{-1}$ in a magnetic field of 4.8 T [19]. Li *et al.,* reported that a small amount doping of Zn for Sn can show a notable change in the type of magnetic transition and the transition temperature [21]. Most recently, Gaonka *et al.,* reported that carbon concentration plays the important role in magnetic properties of $Mn_3SnC$. Although, carbon is non-magnetic (diamagnetic) in nature, its concentration amount in $Mn_3SnC$ can affect the size of $Mn_6C$ unit cell and $Mn_3SnC_{0.80}$, $Mn_3SnC_{0.85}$, $Mn_3SnC_{0.90}$ and $Mn_3SnC_{0.95}$ showed magnetic transition at different transitions at 266 K, 277 K, 283 K and 286 K respectively [18,22-23].

In this paper, we present detailed magneto-transport and thermoelectric studies of $Mn_3SnC$. We observed large non-saturated positive MR in high temperature regime and negative at low temperatures. MR show quadratic dependency on the field in the PM state, whereas, linear dependence in the concurrent AFM/FM state. Both electrons and holes type of charge carriers were observed from Hall resistivity which further confirmed by Seebeck data analysis. We performed Seebeck and Nernst effect in both PM and concurrent AFM/FM state which suggest that both AFM and FM magnons contribute in Seebeck and Nernst signal. Additionally, we calculated a dimensionless parameter which link Seebeck coefficient and electronic specific heat using temperature-dependent Seebeck and Nernst data in the zero-temperature limit within the single band picture for Fermi liquid metals.

**EXPERIMENTAL DETAILS**

Polycrystalline samples of $Mn_3SnC$ were prepared via solid state chemical reaction of the elemental Mn pieces (99.99%), Sn powder (99.99%) and carbon powder (99.9%). The starting materials were mixed in stoichiometric ratio and heated at 800 $^0C$ for 168 hours, in an evacuated quartz tube (pressure ~ $10^{-4}$ mbar). The reacted material was grounded, pelletized and heated again at 800 $^0C$ for 168 hours. The crystal structure was determined by room temperature X-Ray diffraction (XRD) performed via Rigaku Smartlab X-Ray Diffractometer and XRD data was analysed by Rietveld refinement software. The electrical resistivity and Hall effect measurements were performed by standard four-probe method in the temperature range 2 – 350 K and magnetic field up to 8 Tesla by Quantum Design

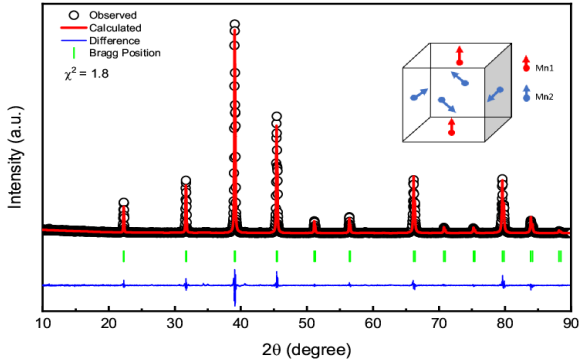

Figure 1. Powder x–ray diffraction pattern showing the clean single phase (*Pm-3m*). Inset in the figure show the orientation of Mn spins in the unit cell. Inset figure shows the magnetic spin alignment of Mn atoms in a unit cell of $Mn_3SnC$ at low temperature.

Physical Property Measurement System (PPMS). The thermal transport (Seebeck and Nernst coefficients) measurements were performed using a homebuilt set-up integrated with PPMS using multifunctional probe assembly [53]. The magnetization measurements were performed by MPMS (SQUID magnetometer). For the electrical and thermal transport measurement the sample were cut into the bar shape geometry.

**RESULTS AND DISCUSSION**

The Reitveld refinement of X-Ray Diffraction pattern shown in Figure 1, revealed that the compound is formed in a single phase cubic structure (space group: *Pm-3m*), with the lattice parameters a = 3.994 Å, which is comparable to the reported values in literature [22-23]. The stoichiometry of carbon atoms plays a very important role in the size of unit cell and properties of the compound. Though it is difficult to determine the atomic percentage of carbon using EDX, the obtained magnetic transition showed our $Mn_3SnC$ to have complete stoichiometry, as reported in the literature [22-23]. The Mn, Sn and C atoms in $Mn_3SnC$ are located at the Wyckoff position 3c (0, 0.5, 0.5), 1a (0, 0, 0) and 1b (0.5, 0.5, 0.5) respectively.

Figure 2a shows, magnetization (M) as a function of temperature measured at 0.01 T magnetic field with a zero field-cooling (ZFC), field-cooling (FC) protocols. A clear magnetic transition is observed around 286 K, where the M (*T*) decrease abruptly corresponding to a transition from high temperature PM state to a low temperature complex magnetic ordering state [18-19]. Inset of figure 2a shows isothermal magnetization M (*H*) at 5 K and it is found that magnetization saturate above 8 Tesla field, with the saturation magnetic moment of ~ 0.38 $\mu_B$/Mn. This smaller magnetic moment of Mn atom a strong itinerant character of d-electrons of the Mn atoms in $Mn_3SnC$ [14].

Figure 2b shows temperature dependence of the longitudinal electrical resistivity $\rho$ (*T*) measured at 0, 4, and 8 T fields. The $\rho$ (*T*) shows sharp drop around 286 K, which indicates the transition from the high-resistive AFM/FM phase to low-resistive PM phase [18-19]. It is clearly seen from figure 2b that resistivity decreases with the increase in the magnetic field, whereas $T_C$ values shift towards the high temperature regime. This is associated with the delocalization of charge carriers caused by the applied magnetic field, resulting in the reduction of the resistivity caused by local ordering of magnetic spins. The AFM/FM transition temperature $T_C$ shifts towards the high temperature regime with application of magnetic field [24-25]. Inset of figure 2b shows the electron-magnon scattering contribution in resistivity. We have fitted the data in the temperature range *T* = 2 – 150 K using the expression;

$$\rho (T) = \rho_0 + \rho_{mag} = \rho_0 + aT^{3/2} + bT^{5/3} \quad .....(1)$$

Where, $\rho_0$ represent the residual resistivity and $T^{3/2}$, and $T^{5/3}$ terms represent the contribution from the AFM and FM magnon scattering respectively. The coefficients of these terms vary on the application of field, which are shown in the Table 1.

Table 1. Magnon scattering parameters

| Parameter | H = 0 | 4 T | 8 T |
|---|---|---|---|
| $\rho_0$ (*$10^{-4}$ mΩ cm*) | 8.17 | 7.96 | 7.74 |
| a (*$10^{-8}$ mΩ-cm.$T^{-3/2}$*) | 20.5 | 6.6 | 0.07 |
| b (*$10^{-7}$ mΩ-cm.$T^{-5/3}$*) | 0.87 | 1.51 | 1.78 |

With the increase in magnetic field, the contribution of AFM magnon scattering decreases whereas the FM magnons contribution increases slightly. It is to mention that resistivity depends on electron-phonon scattering ($\rho$ (*T*) $\propto$ *T*) in the paramagnetic region [26-27]. Moriya, Takimoto and Lonzarich proposed a spin fluctuation mechanism in which AFM-magnon follow $T^{3/2}$ dependency and FM-magnon follow $T^{5/3}$ dependency [28-31]. Thus, the electrical resistivity of

Mn$_3$SnC can be well described by the spin fluctuation induced magnon scattering. Figure 1c shows, the thermal hysteresis in $\rho(T)$ which indicates the first order nature of the transition. The inset of figure 2(c) shows the decrease in width of hysteresis with the field.

The transverse magnetoresistance (MR) of Mn$_3$SnC is measured in the temperature range 2 – 300 K under magnetic field up to 8 T as shown in figure 3a, which is defined as MR = [ρ (H) - ρ (0)]/ρ(0) × 100%. We observed two distinct features in the MR data. First, a non-saturating positive MR value, and secondly, both positive and negative MR values are obtained with in different temperature regime. These features are seen more clearly in contour plot (Figure 3a). From this figure, it is quite noticeable that in the concurrent AFM/FM state, MR monotonically decreases as temperature decreases and become negative at lower temperatures. MR exhibits sharp increase at the boundary of magnetic transition. The maximum MR value is found as ~ 8.2% for 8 T at 286 K. The positive

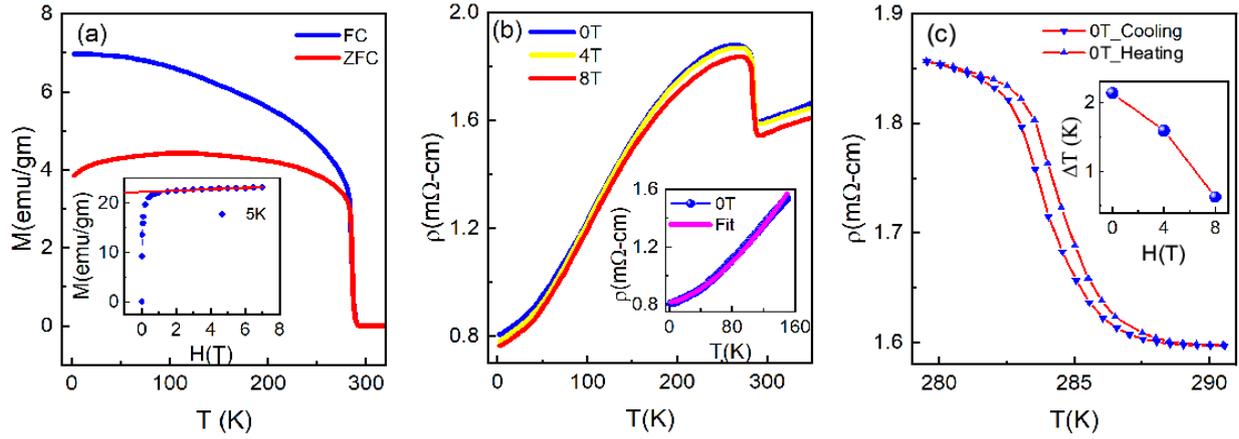

Figure 2.(a) Magnetization as a function of temperature under ZFC and FC process at H = 0.01 T. Inset shows the isotherm magnetization M(H) at T = 5 K. (b) The $\rho(T)$ vs T at H = 0, 4, and 8 T and inset shows the electron-magnon scattering dependency fitting using equation (1). (c) Thermal hysteresis around transition temperature. Inset represents the dependency of thermal hysteresis on magnetic field

value of MR is quite unusual in these types of system because the application of field suppresses the randomness of spins, which should lead to a negative MR. Furthermore, non-saturating MR is usually predicted for YPdSi and GdPdSi semimetals [32]. The figure 3a depicts a weak thermal hysteresis width of about 0.15 and 0.12 T at the temperature 286 K and 288 K respectively in the 5 T magnetic field [33]. The hysteresis width decreases with the increase temperature and no hysteresis is observed above $T_C$. MR data can be good fitted with the power law MR $\alpha H^n$. Parish-Littlewood (PL) [34-36] proposed a model that inhomogeneity in the carrier density can produce a linear MR in the presence of hall field. The non-saturating linear MR in Mn$_3$SnC can also be ascribed due to the inhomogeneity as suggested by PL model. The inset of figure 3b shows variation of power factor $n$ with the temperature. Around the transition temperature, $n$ value is estimated as ~1.83 which indicates the compensated electrons and hole densities, predicted by two band model theory [30].

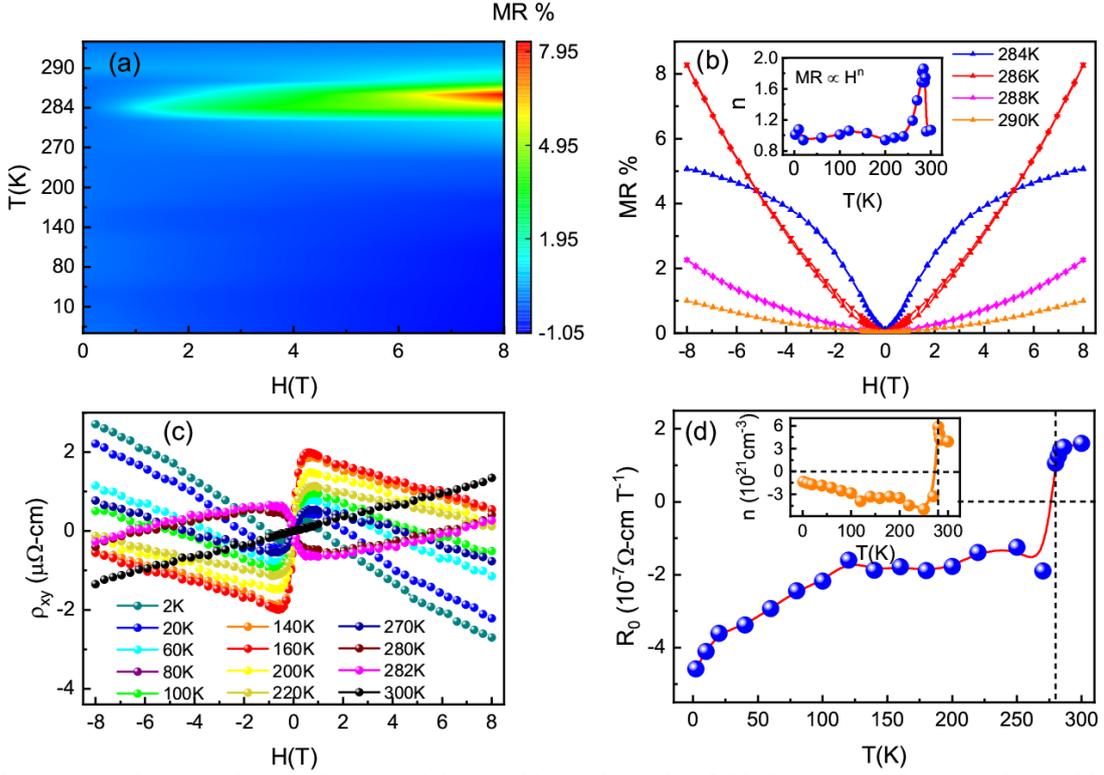

Figure 3. (a) MR% shows maximum value around the transition and H = 8 T field. (b) MR near the magnetic transition (c) Transverse resistivity near the transition showing the change at the transition. (d) Hall constant $R_H$ versus $T$, shows a cross over from hole to electron dominating conduction and inset of figure 3(d) shows the carrier density as a function of temperature.

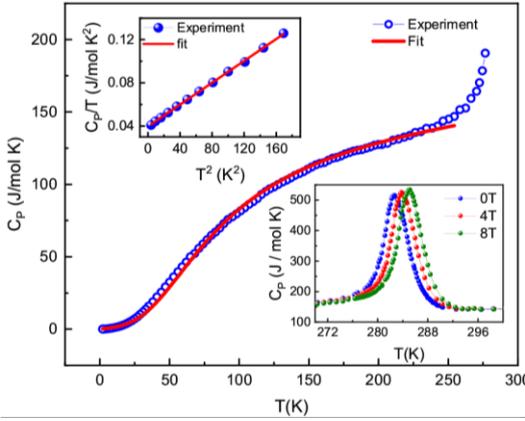

Figure 4. Temperature dependence of specific heat capacity. Upper and lower inset of figure 4. are showing the fitted $C/T$ vs $T^2$ plot and field dependent $C_P$ near the transition temperature for the same.

and hole densities, predicted by two band model theory [37].

The Hall effect measurements were performed for a temperature range of 2 – 300 K up to 8 T field. The experimental data of Hall resistivity were anti-symmetrized to remove the contribution from the longitudinal resistivity using the formula $[\rho_{xy}(H) - \rho_{xy}(-H)]/2$ as shown in figure 3(c). The total Hall resistivity $\rho_{xy}(H)$ of Mn$_3$SnC has two types of contributions, such as normal Hall and anomalous Hall, and it can be written as $\rho_{xy}(H) = R_0H + R_SM_S$, where $R_0$, $R_S$, $H$ and $M_S$ represent the ordinary, anomalous Hall coefficients, external applied field and saturation magnetization respectively. The $\rho_{xy}(H)$ curves increase sharply at low field and show linear dependence in the high field region, which indicates the contribution of an anomalous signal in Hall effect measurements. $\rho_{xy}(H)$ increases with increasing temperature, attain maximum value at 160 K and decreases with further increasing temperature. Near transition temperature, field dependence is changed, and $\rho_{xy}(H)$ shows linear field dependence at 300 K, indicating a normal Hall effect. To separate out the normal and anomalous Hall contribution, we performed linear fitting of the high field region of $\rho_{xy}(H)$. The slope of the linear fit on the y-axis gives $R_0$ [38-39]. We show the variation of $R_0$ with the temperature in figure 3(d); the negative and positive signs of $R_0$ indicate the presence of electron-like and hole-like charge carriers in concurrent AFM/FM states and PM state respectively. The inset of figure 3(d). shows carrier density $n$ as a function of temperature, calculated using the relation $R_0 = 1/ne$, where $e$ is the electronic charge. At lowest temperature, the n value is found to be ~ $3.9 \times 10^{21}$, and the corresponding mobility ($\mu$) is ~ 15.6 cm$^2$ V$^{-1}$ S$^{-1}$.

We measured the heat capacity ($C_p$) in the temperature range 1.8 – 330 K at 0, 4 and 8 T magnetic field (Figure 3). A sharp spike anomaly near $T \sim 284$ K is observed, which confirms the FM ordering at $T_C$ showed in our magnetization measurements. Data is fitted by following expression which has combination of linear electronic contribution and phonon contribution of Debye type in the temperature range 1.8 – 200 K [40].

$$C_P(T) = \gamma T + 9nR \left(\frac{T}{\theta_D}\right)^3 \int_0^{\theta_D/T} \frac{x^4 e^x}{(e^x-1)^2} dx \quad ..(2)$$

where, $\gamma$ is known as Sommerfeld coefficient, n is the number of atoms in the compound, R is the gas constant and $\theta_D$ is the Debye temperature. The obtained values of $\gamma$ and $\theta_D$ from the above expression are ~ 39.01 mJ/mol-$K^2$ and 308 K. The low temperature part of the $C/T$ vs $T^2$ is plotted in upper inset of figure 3. The $\gamma$ value obtained is estimated to be ~ 36.04 mJ/mol-$K^2$, and the $\beta$ value is estimated to be 0.5 mJ/mol $K^4$. We have shown the field dependence of heat capacity near $T_C$ in the in lower inset of figure 3. On increasing the magnetic field, peaks shift towards the high temperature side which is consistent with the electrical resistivity data.

Temperature dependence of Seebeck coefficient ($S = S_{xx} = \Delta V_{xx}/\Delta T_{xx}$) is shown in figure 5a. $S$ measured in the temperature range from 2 – 340 K under 0, 2, 4, 6 and 8 T magnetic fields. $S(T)$ shows abrupt change at transition temperature $T_C \sim 286$ K as we observed in other transport and magnetic properties. The sign of $S(T)$ is consistent with the Hall effect measurement; positive in PM state and negative in concurrent AFM/FM state. In general, $S$ is defined as the sum of diffusion ($S_d$) term and the phonon drag ($S_P$) terms; but phonon drag term is important at $T_{max} \sim \theta_D/20$ [41]. Gurvich's theory suggested that the linear dependency of applied magnetic field on $S$ is related to phonon drag contribution at low temperature [42-43]. However, the $S(T)$ of $Mn_3SnC$ does not show linear dependency with the applied field at the low-temperature ($T \sim 50$ K) as shown in figure 5f, which implies that phonon drag contribution can be neglected. In polycrystalline samples, it is difficult to see a large contribution from the phonon drag due to the presence of large amount crystal defects. Similar to the phonon-drag effect, electron-magnon scattering produce magnon drag effect which might be a good candidate in present case. Blatt et al. [44] predicted magnon drag contribution in iron by applying the magnon drag theory, similar to the phonon drag theory. Experimentally, the magnon drag contribution was first explained in magnetic systems Fe, Co and Ni with the sum of diffusion contribution [45]. As $Mn_3SnC$ has a concurrent AFM/FM state below 286 K, scattering will be dominated by both AFM and FM-magnons. In the presence of electron-magnon scattering mechanism, $S(T)$ is expected to follow $T^3$ dependency for AFM-magnon contribution and as $T^{3/2}$ for FM-magnon contribution [43,46-48]. In addition, $S(T)$ may also contain electron-diffusive and spin wave fluctuation contributions. Thus, $S(T)$ contrition for magnetic systems can be given by the expression;

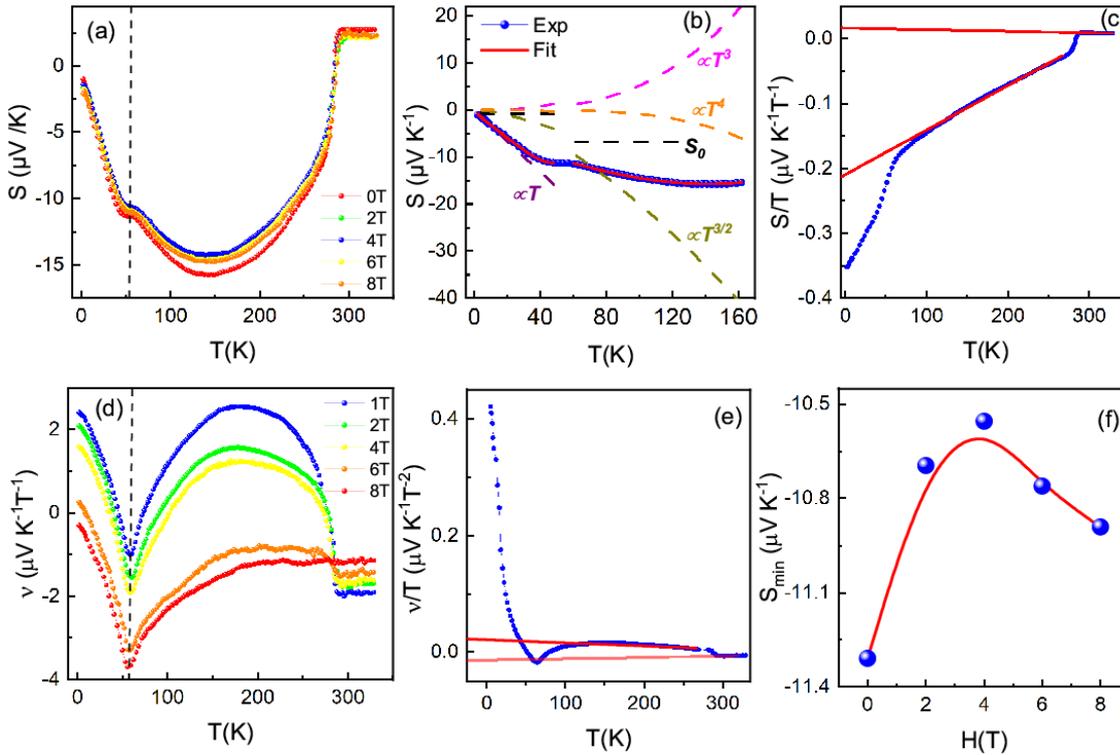

Figure 5. (a) Temperature dependence of Seebeck coefficients at various constant magnetic fields. (b) shows the diffusive, magnon and spin wave fluctuation fitting and red line correspond to the equation 3. (c) shows the $S/T$ and red lines represent the zero-temperature limit in both concurrent AFM/FM and PM state. (d) temperature dependence of Nernst coefficient at various constant magnetic fields. (e) shows the $\nu/T$ and red lines represent the zero-temperature limit in both concurrent AFM/FM and PM state. (f) present field dependence of Seebeck coefficient at $T \sim 50$ K.

$$S(T) = S_0 + S_1 T + S_{3/2} T^{3/2} + S_3 T^3 + S_4 T^4 \quad \ldots\ldots(3)$$

where, $S_0$ is a constant term, $S_1 T$ represents diffusive contribution, $S_{3/2} T^{3/2}$ represent FM-magnon contribution, $S_3 T^3$ represent the AFM-magnon contribution and $S_4 T^4$ represent the spin wave fluctuations in the magnetically order state. As shown in the Fig 5 (b), we have two distinct behaviours below 160 K, separated by an anomaly at ~ 50 K which might either be associated with the magnon drag or the restructuring of the Fermi surface. Therefore, we have fitted the data in two separate regions $T = 2 - 50$ K and $60 - 160$ K using equation (3). For $50 < T < 160$ K, $S(T)$ fitting do not require the inclusion of the electron diffusive term. The corresponding fitting parameters for both the temperature ranges are given in table 2. Further we have plotted the contributions of each term in figure 5b, that clearly show the range and strength of AFM magnon, FM magnon, and the spin fluctuation driven term in the total $S(T)$. It is to mention that the correlated systems, positive AFM magnon term is arising out of the scattering of the magnon with the minority carriers (holes). It is clear that spin wave fluctuation has a very small contribution in both regions. Another highlight is the formation of a broad minimum in $S(T)$ under absence and presence of magnetic field at high temperature and the corresponding valley position does not change with applied magnetic fields. The upturn in $S(T)$ around $T \sim 150$ K is thought to be caused by the thermal excitation of quasiparticles across their pseudo gap. Similar feature has been observed in Fe$_2$VAl, Fe$_2$TiSn and Fe$_2$VGa$_{1-x}$Ge$_x$ semimetal Heusler alloys [49-51]. In Mn$_3$SnC, Fermi level is found above the pseudo gap and corresponding a large negative value is obtained in $S(T)$ without modification in electronic structure. On the other hand, it was observed that there is no irregularity in $\rho(T)$.

Table 2. Various fitting parameters after fitting to the 2 – 160 K temperature Seebeck data.

| Parameters | $T = 2 – 50$ K | $T = 60 –160$ K | | | | |
|---|---|---|---|---|---|---|
| H | 0 - 8 T | 0 | 2 T | 4 T | 6 T | 8 T |
| $S_0$ (µV K$^{-1}$) | -0.350 | -7.16 | 6.50 | -6.15 | -6.79 | -6.70 |
| $S_1$ (µV K$^{-2}$) | -0.264 | --- | --- | --- | --- | --- |
| $S_{3/2}$ ($10^{-2}$ µV K$^{-1.5}$) | -0.012 | -1.10 | 1.14 | -1.20 | -1.11 | -1.20 |
| $S_3$ ($10^{-6}$ µV K$^{-4}$) | 0.61 | 4.22 | 5.93 | 6.53 | 5.38 | 6.60 |
| $S_4$ ($10^{-9}$ µV K$^{-5}$) | -0.002 | -5.12 | -13.55 | -15.73 | -11.0 | -16.07 |

Diffusive contribution of Seebeck signal is expected to be $T$-linear in zero-temperature limit, with a magnitude proportional to the strength of electronic correlation as in the case of the $T$-linear electronic specific heat, C$_e$/T = $\gamma$. Both relations can be related through Fermi temperature $T_F$. In metallic system, Seebeck coefficient is given by

$$S(T) = \pm \frac{\pi^2 K_B}{3e} T \left( \frac{d \ln D_\pm(E)}{dE} \right) \quad \ldots (4)$$

$$\frac{S}{T} = \pm \frac{\pi^2}{2} \frac{k_B}{e} \frac{1}{T_F} \quad \ldots (5)$$

Where, D(E$_+$) and D(E$_-$) represent density of states (DOS) of holes and electrons respectively, $k_B$ is the Boltzmann constant. We plotted $S/T$ vs $T$ in figure 5c which goes from positive at PM state to negative at concurrent AFM/FM state, in agreement with a similar sign change $R_H$. We calculated Fermi energy ($E_F$), Fermi temperature ($T_K$), Fermi radius ($k_F$), effective mass (m*), Fermi velocity ($v_F$) and carrier concentration in zero temperature limit in both magnetic states. Using free-electron gas approximation, carrier concentration ($n$) can be calculated using following relation

$$E_F = \frac{(hc)^2}{8 m_e c^2} \left( \frac{3}{\pi} \right)^{2/3} n^{2/3} \quad \ldots (6)$$

where $h$ is Planck's constant, c is velocity of light, and m$_e$ is mass of electron. The value of n is estimated n ~ $1.44 \times 10^{21}$ cm$^{-3}$ which is very close to our Hall effect result. The density of states $D(E_F)$ at the Fermi level is estimated as $D(E_F) = 2.69$ states eV$^{-1}$ f.u$^{-1}$ from the Sommerfeld coefficient $\gamma$ using the expression; $\gamma = \frac{\pi^2}{2} k_B \frac{n}{T_F} = \frac{\pi^2}{3} k_B^2 D(E_F)$, where R represents the ideal gas constant, and $n$ is number of atoms in Mn$_3$SnC. Behnia et al. [52] introduced a dimensionless parameter (q) which is related to the electronic specific heat ($\gamma$) and $S(T)$ is given by $q = \frac{S(T)}{T} \frac{N_{av} e}{\gamma}$, where $N_{av}$ represent the Avogadro number. Assuming single band transport and a spherical surface in concurrent

Table 3. Parameters obtained from zero temperature limit of Seebeck and Nernst data within the single band picture in Fermi liquid

| Parameters | PM state (Hole type) | AFM/FM state (Electron type) |
|---|---|---|
| $S/T$ (μV K$^{-2}$) | 0.0107 | -0.24 |
| $\nu/T$ (μV K$^{-2}$ T$^{-1}$) | -0.0140 | 0.023 |
| $E_F$ (eV) | 3.36 | 0.152 |
| $T_F$ (10$^3$ K) | 39.0 | 1.76 |
| $\mu$ (cm$^{-2}$ V$^{-1}$s$^{-1}$) | 23.20 | --- |
| $n$ (10$^{21}$ cm$^{-3}$) | 27.8 | 3.28 |
| $q$ | 0.032 | -0.72 |
| $k_F$ (10$^7$ cm$^{-1}$) | 12.2 | 6.24 |
| $m^*$ | 1.3 m$_e$ | 0.31 m$_e$ |
| $v_F$ (10$^6$ m/sec) | 1.05 | 23.1 |

AFM/FM state, we have obtained Fermi wave vector ($k_F$), effective mass ($m^*$) and Fermi velocity ($v_F$) given by $k_B T_F = \hbar k_F^2/2m*$. Earlier, the value of q (~ 0.8) has been observed for Sr$_2$RuO$_4$ as we observed in Mn$_3$SnC in its PM state due to the same γ value [52].

The Nernst effect is the transverse electric field produced by a longitudinal thermal gradient in the presence of a magnetic field. Figure 5d shows Nernst signal ($\nu = S_{xy}/H; S_{xy} = \triangle V_{xy}/\triangle T_{xx}$) measured in the temperature range 2 – 340 K in the presence of 1, 2, 4, 6 and 8 T magnetic fields. The Nernst effect offers a way to understand properties of topological materials. The Nernst signal provides information on small unusual contributions, including extrinsic contributions from magnetic impurity scattering and intrinsic contributions from Berry curvature [54-55]. Around $T_C$, spin fluctuations have more significant impact, resulting in Nernst signal exhibits drastic change with the field. A clear PM to concurrent AFM/FM transition is observed for 1 T field, however as field increases, this magnetic transition gets suppressed and almost disappears for 8 T field. The suppression of spin fluctuations under external fields helps to explain this crossover [47]. In PM state, Nernst signal moves upward with increasing the field, while in AFM/FM state move downward which implies that spin's ordering increase as magnetic field increase, as a result movement of carriers become slow. The Nernst signal show a sharp drop at $T \sim 50$ K, which may be due to the re-structuring of Fermi surface in this temperature range. It is found that sharp drop is independent of the magnetic field. The existence of Dirac dispersion bands at the Fermi level can produce large value of Nernst signal [54]. The Nernst signal of Bi$_2$Se$_3$ is proposed 2.3 μV K$^{-1}$T$^{-1}$ [54] and Mn$_3$SnC show maximum value ~ 3.5 μV K$^{-1}$T$^{-1}$ around 50 K at 8 T field. It is to note that Nernst coefficient can be of any sign, without any direct relation to carrier type. The $\nu$ ($T$) is related to the Fermi temperature $T_F$ as,

$$\nu(T) = \frac{\pi^2}{3} \frac{K_B}{e} \frac{T}{T_F} \mu \quad …(7)$$

Where, k$_B$ is the Boltzmann constant, e is the electron charge and $\mu$ is the carriers mobility. This formula relates Nernst signal to charge carrier mobility. This formula relates the Nernst signal with the carrier mobility. The values of $\nu/T$ are obtained -0.0140 μV K$^{-2}$ T$^{-1}$ and 0.023 μV K$^{-2}$ T$^{-1}$ in the zero-temperature limit for the PM state and concurrent AFM/FM respectively, as shown in figure 5e. From the above expression low value of carrier mobility (~23.2 cm$^2$ V$^{-1}$ S$^{-1}$ at 2 K) is obtained which is very close to our Hall result. The obtained value of carrier mobility is similar to Nb$_{0.20}$Bi$_2$Se$_3$ topological superconductor [32]. Table 3 summarizes the parameters obtained from the zero-temperature limit of thermopower.

## CONCLUSIONS

In summary, we have investigated the low-temperature magneto-transport and thermal transport properties. It is found that Mn$_3$SnC undergo a magnetic transition from PM to concurrent AFM/FM state at $T_C$ ~ 286 K temperature. The low temperature electrical resistivity and thermopower data (Seebeck and Nernst effect) show an electron-magnon scattering dependency. Heat capacity show a peak at $T_C$ which shifts towards high temperature as magnetic field increases. The MR and Seebeck signal show exhibit topological semimetal properties. Another highlight of thermopower is that a pseudo gap near the Fermi level produces an excess of charge carriers, resulting in a large signal of thermopower. MR is found positive and reaches ~ 8.5% at 8 T in the paramagnetic region and negative in concurrent AFM/FM state at low temperature. It is found that Mn$_3$SnC show multiband character from the Hall resistivity and Seebeck effect studies. Furthermore, the presence of thermal hysteresis in resistivity and magnetoresistance around magnetic transition suggests that system goes first order magnetic transition. The ARPES studied on single crystalline Mn$_3$SnC would be useful to elucidate Weyl nodes near the Fermi level and the possible re-structuring of the Fermi surface near 50 K.

## ACKNOWLEDGEMENTS

We acknowledge Advanced Material Research Center (AMRC), IIT Mandi for the experimental facilities. SG, and Sonika acknowledge IIT Mandi and MoE, India for the HTRA fellowship. CSY acknowledges SERB-DST (India) for the CRG grant (CRG/2021/002743).

## DATA AVAILABILITY

The datasets generated during and/or analyzed during the current study are available from the corresponding author on reasonable request.